%% file: main.tex
\documentclass{article}
\usepackage{spconf,amsmath,graphicx}
\usepackage{wasysym}
\usepackage{booktabs}
\usepackage{commath} 
\usepackage{mathtools} 
\usepackage{amssymb}
\usepackage{caption}
\usepackage{subcaption}
\usepackage{hyperref}
\usepackage{nicefrac}
\usepackage{bm}
\usepackage{multirow}
\usepackage{calrsfs}

\DeclareMathAlphabet{\pazocal}{OMS}{zplm}{m}{n}

\title{Meta-learning Extractors for Music Source Separation}

\twoauthors
 {David Samuel\sthanks{Work completed during an internship at Preferred Networks.}}{Charles University\\
	 Prague}
 {Aditya Ganeshan, Jason Naradowsky}
	{Preferred Networks\\
	 Tokyo}

\begin{document}

\maketitle

\input{00_abstract.tex}
\input{01_intro.tex}

\input{02_method.tex}
\input{03_improvements.tex}
\input{04_experiments.tex}
\input{05_related.tex}
\input{06_conclusion.tex}

\bibliographystyle{IEEEbib}
\bibliography{strings,refs}

\end{document}

%% file: 00_abstract.tex
\begin{abstract}
We propose a hierarchical meta-learning-inspired model for music source separation (Meta-TasNet) in which a generator model is used to predict the weights of individual extractor models. This enables efficient parameter-sharing, while still allowing for instrument-specific parameterization.  Meta-TasNet is shown to be more effective than the models trained independently or in a multi-task setting, and achieve performance comparable with state-of-the-art methods.  In comparison to the latter, our extractors contain fewer parameters and have faster run-time performance. We discuss important architectural considerations, and explore the costs and benefits of this approach.
\end{abstract}


%% file: 01_intro.tex
\section{Introduction}
\label{sec:intro}

Mankind's enduring and nearly universal appreciation of music has inspired the creation of thousand of instruments, each with its own unique timbral qualities.  Yet there are strong similarities in the sonic characteristics of many instruments.  A saxophone and a clarinet utilize similar methods for producing sound and thus exhibit similar timbral characteristics across time.  A soprano singer and trumpet may differ categorically, but occupy similar frequency bands.  If our models are aware of such relationships, there is potential to tailor specific separation strategies to each instrument, while making better use of the training data as a whole.

In this work we explore the application of ideas from meta-learning and AutoML to the problem of source separation.  Our goal is to generate instrument-specific high-precision separation models, each of which is finely-tuned for dealing with the nuances of a particular instrument.  However, rather than train each of these models directly, we train a separate generator network to predict their parameters.  Thus the generator network is able to understand the relationships between instruments, and take them into account when generating specific separation networks (here, the masking subnetwork of a ConvTasNet~\cite{conv-tasnet}).  This functions as a form of parameter sharing, allowing the training data for one instrument to benefit another.  The resulting extractors achieve greater performance with fewer parameters.

Our contributions are the following:
\begin{enumerate}
    \itemsep-0.08em 
    \item To our knowledge we are the first to apply the network-generating network approach to the problem of source separation, where we show it outperforms naive training of instrument-specific separator networks.
    \item In comparison to a single multi-task model, our models perform better, and are smaller and faster.
    \item We describe a number of improvements for Conv-TasNet, and our final architecture achieves state-of-the-art performance on a number of MUSDB18 tasks, a first for waveform-based separation models.
\end{enumerate}

%% file: 02_method.tex
\section{Generating Extractor Models}

\begin{figure}
\centering
\includegraphics[width=0.9\columnwidth]{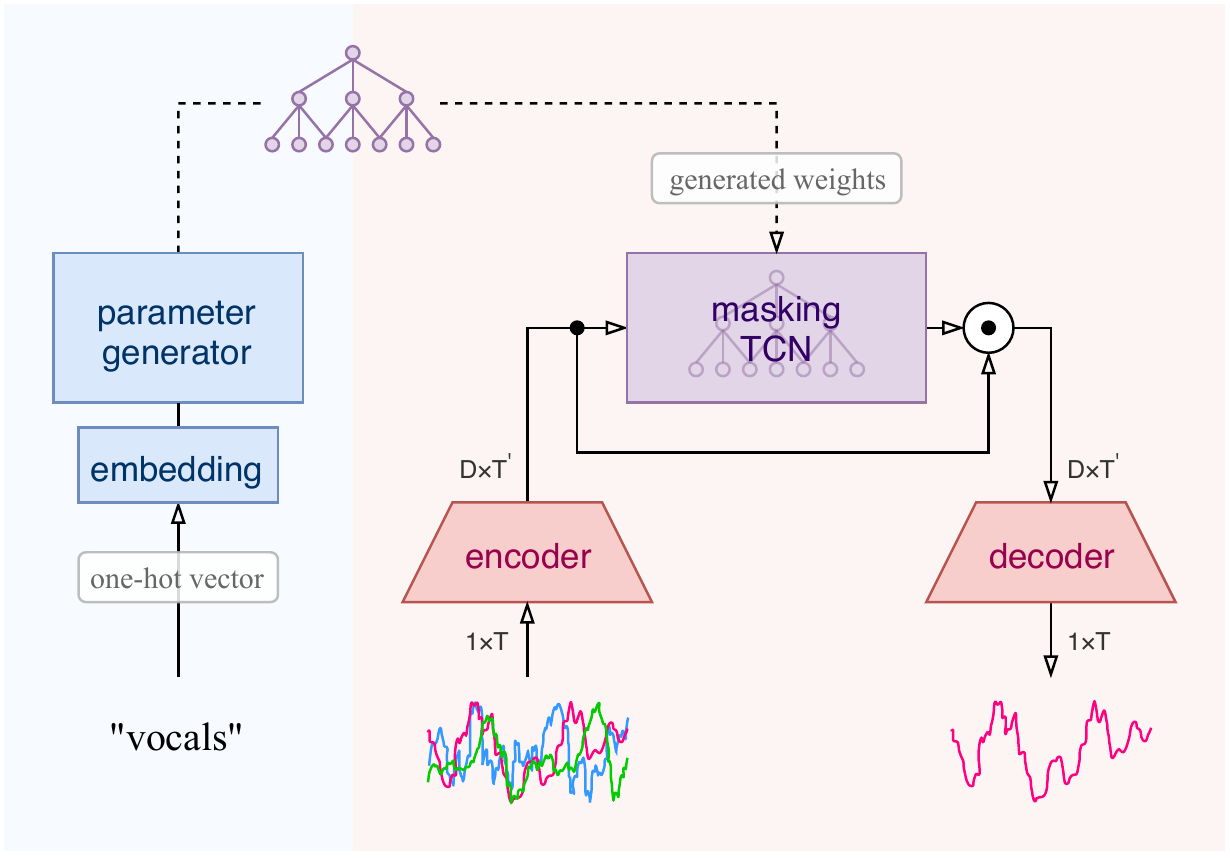}
\caption{The overall architecture. The blue area depicts the parameter generator, a network which predicts the weights of the extractor's masking subnetwork specific to each instrument.  The extractor network then uses these weights when separating the instrument source from the mixture.}
\end{figure}

The key idea is to utilize a tiered architecture where a \emph{generator} network ``supervises'' the training of the individual extractors by generating some of their parameters directly.  This allows the generator to develop a dense representation of how instruments relate to each other \emph{as it pertains to the task}, and to utilize their commonalities when generating each extractor.

\subsection{Extractor Model}
\label{sec:extractor}

Our model is based on Conv-TasNet~\cite{conv-tasnet}, a time domain-based approach to speech separation comprising three parts: (1) an encoder which applies a 1-D convolutional transform to a segment of the mixture waveform to produce a high-dimensional representation, (2) a masking function which calculates a multiplicative function which identifies a targeted area in the learned representation, and (3) a decoder (1-D inverse convolutional layer) which reconstructs the separated waveform for the target source.  The use of an intermediate representation overcomes the difficulty of working with high resolution in the time dimension (tens of thousands samples per second) without the disadvantages of compressing the data via an unlearned transformation, such as a mel spectrogram.

Given an input mixture \(\mathbf{s} = \sum_{i \in \mathrm{I}}\mathbf{s}_i \in  \mathbb{R}^{T}\) of all the sources \(\mathrm{I}\), the encoder maps it into a latent representation \(\mathbf{h} = \mathrm{encoder(}\mathbf{s}) \in \mathbb{R}^{D\times T'}\) via a 1-D convolution. 
This is forwarded to the masking subnetwork, a temporal convolutional network (TCN), whose outputs are the separation masks \(\mathbf{m}_i = \mathrm{mask_i(}\mathbf{h}) \in [0,1]^{D \times T'}\). The separated latents \(\mathbf{\bar{h}}_i\) are then obtained by the element-wise product \(\mathbf{\bar{h}}_i = \mathbf{h} \odot \mathbf{m}_i\). After masking, the final extracted signal \(\mathbf{\bar{s}}_i\) is returned by a transposed 1-D convolution of the decoder: \(\mathbf{\bar{s}}_i = \mathrm{decoder(}\mathbf{\bar{h}}_i) \in \mathbb{R}^{T}\).

The masking network is of particular interest, as it contains the source-specific masking information; the encoder and decoder are source-agnostic and remain the same for separation of all sources.

\subsection{Meta-learning Extractor Parameters}

The \emph{generator} is a network that predicts the parameters of a secondary network, the baseline model which we refer to as an \emph{extractor}, conditioned on additional information.  As it pertains to this work, additional information is the identity of the instrument to be separated, provided as a one-hot vector $i$.  This vector is projected into \(\mathbf{e}_i \in \mathbb{R}^{M}\) where the generator can encode the attributes -- similarities and dissimilarities -- of the instruments along multiple axes.

As the encoder and decoder are defined as instrument-agnostic components, we focus solely on using the generator to predict the parameters of the masking subnetwork.  As described in Sec.~\ref{sec:extractor}, this
 consists primarily of a series of TCN layers.  The generator function defines the the weights and biases of the \(k\)-th layer in the extractor network as:
\begin{equation}
    \bm{\theta}_k \vcentcolon = \mathbf{W}_k\mathbf{P}_k\mathbf{e}_i 
\end{equation}

\noindent where \(\mathbf{e}_i \in \mathbb{R}^{M}\) is the learned embedding of an instrument \(i\) and \(\mathbf{P}_k \in \mathbb{R}^{M' \times M}\), \(\mathbf{W}_k \in \mathbb{R}^{|\bm{\theta}_k| \times M'}\) are learnable linear functions. The definition is further constrained with \(M' < M\) so that the mapping by \(\mathbf{P}_k\) extracts the most relevant information from \(\mathbf{e}_i\) which is then transformed with \(\mathbf{W}_k\) to yield all the parameters of the \(k\)-th layer (similar to~\cite{platanios-etal-2018-contextual}).  Other aspects of the model remain unchanged.

%% file: 03_improvements.tex
\section{Adaptations for Time-Domain Music Source Separation}
\label{subsec:improvements}

\begin{figure}
\centering
\hspace{-1em}
\includegraphics[width=0.9\columnwidth]{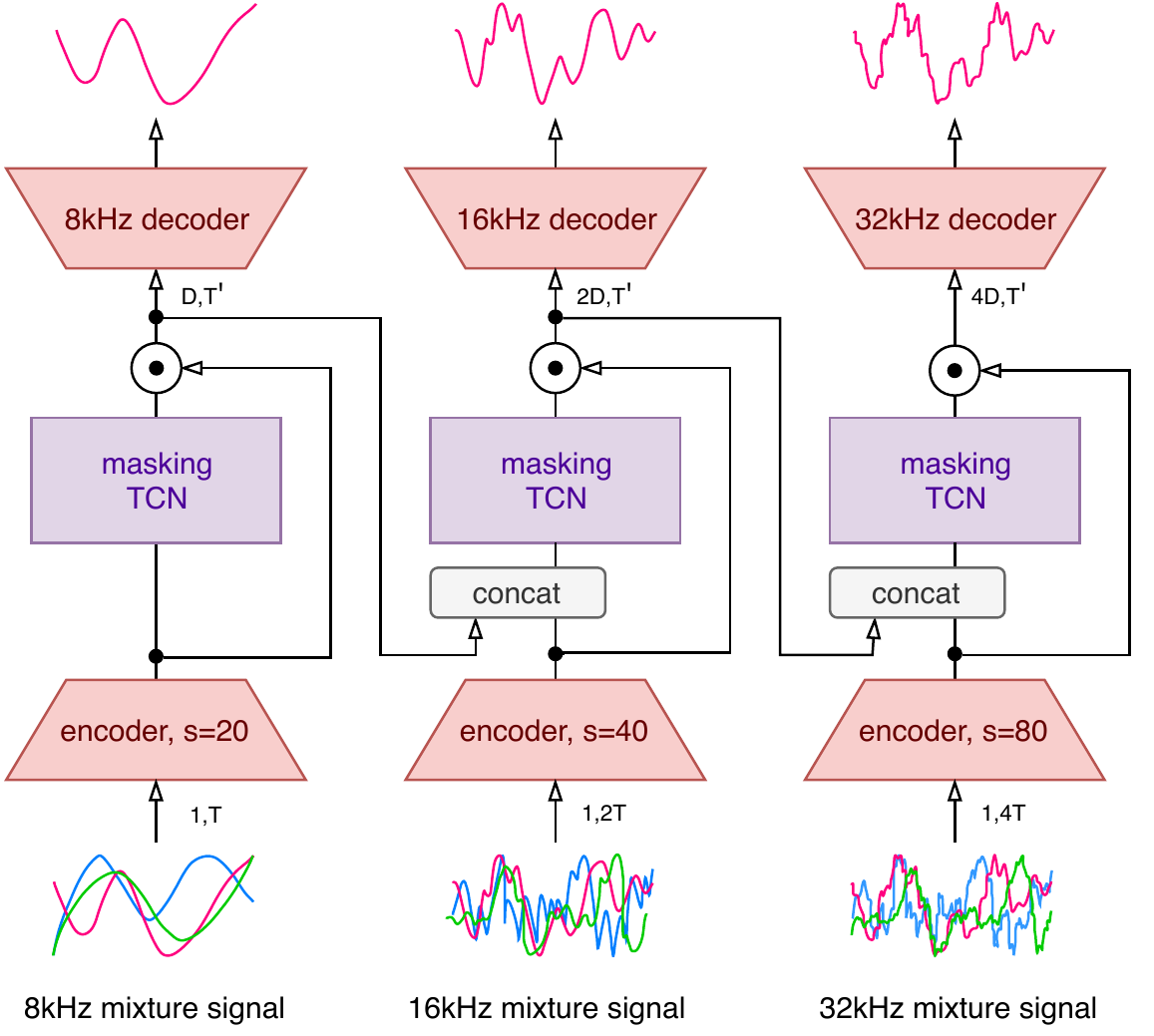}
\caption{Illustration of the multi-stage architecture. The resolution of the estimated signal is progressively enhanced by utilizing information from previous stages. The encoders increase the stride $s$ to preserve the same time dimension $T'$.}
\label{fig:multi_stage}
\end{figure}

Additionally, we observe better performance by modifying the Conv-TasNet architecture as follows:

\subsection{Multi-stage Architecture}

We observe that models trained using lower sampling rates perform better on 44kHz separation, despite the loss in resolution.
We therefore propose a multi-stage architecture (Fig. \ref{fig:multi_stage})
which begins by predicting low resolution audio, and iteratively upsamples at each stage (similar to \cite{laplacian_pyramid_gan}).

After computing a mask \(\mathbf{m}_i\) and applying it on \(\mathbf{h}\), \(\mathbf{\bar{h}}_i\) is forwarded into the next stage (apart from being decoded into the time domain). At each stage the sampling rate is increased and the encoder's kernel width, stride size, and output size are increased proportionally.
Note that this preserves the number of time-steps $T'$ in all stages. 
We use three stages with 8, 16 and 32kHz sampling rates; the mask in the last 32kHz stage is calculated as  \(\mathbf{m}_i^{\mathrm{32}} = \mathrm{mask(concat(}\mathbf{\bar{h}}_i^{\mathrm{16}}, \mathbf{h}^{\mathrm{32}}\mathrm{))}\).

\subsection{Auxiliary Loss Functions}

Working in time-domain allows us to optimize scale-invariant signal-to-noise ratio (SI-SNR) in an end-to-end manner.  This is an approximation of the signal-to-distortion (SDR) ratio metric used for the final evaluation~\cite{SiSEC18}.

We additionally utilize three  auxiliary loss functions for improved training, each a component in a weighted sum which constitutes the final loss. We denote \(\mathbf{s}_{b,i} \in \mathbb{R}^T\) as the ground-truth separated signal for an instrument \(i \in \mathrm{I}\), \(b \in \mathrm{B}\) as the index of \textit{b}th training sample in a batch, and \(\mathbf{h}_{b,i} = \mathrm{encoder}(\mathbf{s}_{b,i}) \in \mathbb{R}^{D \times T'}\) as the separated latent space. The auxiliary losses can now be defined as:   
\begin{enumerate}
    \item \textbf{Dissimilarity Loss}: For each training sample, this loss minimizes the similarity between the different instrument representations $\mathbf{h}_{b,i}$:
        \begin{equation}
            \label{eq:l_diss}
            \hspace{-2em}\pazocal{L}_{\mathrm{diss},b} \vcentcolon= {\binom{\vert \mathrm{I} \vert}{2}}^{-1} \sum_{i\neq j \in \mathrm{I}}\frac{\mathrm{abs}(\mathbf{h}_{b,i})\cdot \mathrm{abs}(\mathbf{h}_{b,j})}{\Vert \mathbf{h}_{b,i} \Vert \Vert \mathbf{h}_{b,j} \Vert}
        \end{equation}
    \item \textbf{Similarity Loss}: For each instrument, this loss maximizes the similarity between instrument representations in different training samples:
        \begin{equation}
            \hspace{-2em}
            \label{eq:l_sim}
            \pazocal{L}_{\mathrm{sim},i} \vcentcolon= - {\binom{\vert \mathrm{B} \vert}{2}}^{-1} \sum_{b\neq b' \in \mathrm{B}}\frac{\mathbf{h}_{b,i}\cdot \mathbf{h}_{b',i}}{\Vert \mathbf{h}_{b,i} \Vert \Vert \mathbf{h}_{b',i} \Vert}
        \end{equation}
    \item \textbf{Reconstruction Loss}: This loss increases the SI-SNR between the mixture signal \(\mathbf{s}\), and signal as processed without any masking, \(\mathbf{\hat{s}} = \mathrm{decoder}(\mathrm{encoder}(\mathbf{s}))\).
\end{enumerate}

\subsection{Stronger Encoder}

\begin{figure}
\centering
\includegraphics[width=0.55\columnwidth]{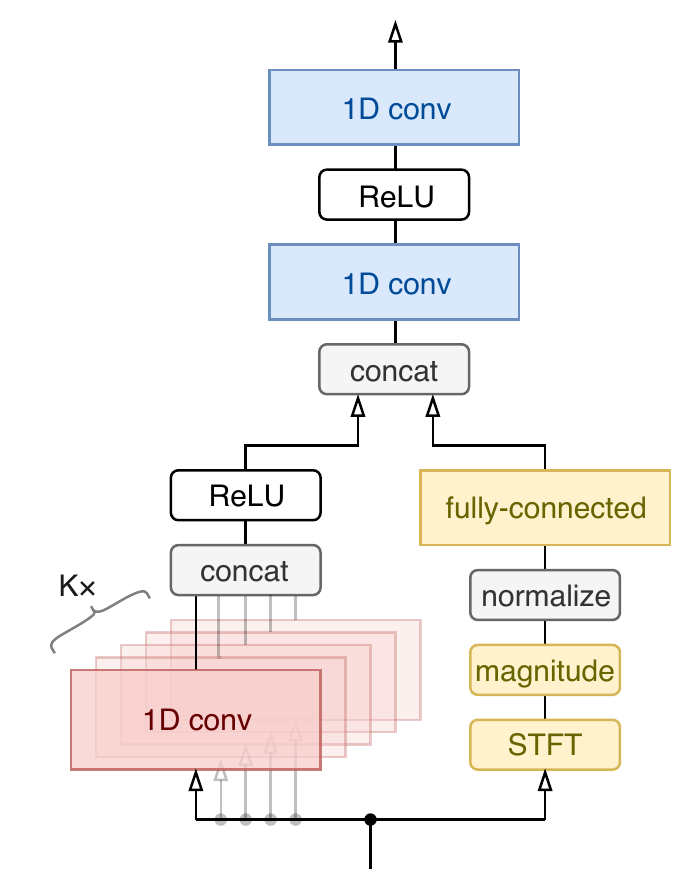}
\caption{Illustration of our encoder architecture. The input (mixture) signal is transformed with multiple convolutional heads (red) and standard SFTF (yellow). These two branches are then merged and mapped onto the latent space (blue) where the outcome is later separated.}
\label{fig:encoder}
\end{figure}

The original formulation of the (Conv-)TasNet uses a single 1-D convolutional layer encoder as a learnable replacement of the STFT. We utilize a more complex encoder capable of capturing more distinct features from the input signal (Fig.~\ref{fig:encoder}).  We concatenate the information gathered by \(K\) 1-D convolutional layers where the \(k\)-th layer has kernel size \(\nicefrac{1}{2^k}W\) and output dimension \(\nicefrac{2^k}{2^K}D\). In this way, the multiple kernels are able to capture a wider frequency range with more fidelity. We also include features from the classical STFT spectrogram of the input mixture, normalizing it, and projecting it down with one linear transformation (as a learnable replacement for a mel filter). These two branches are merged and run through Conv-ReLU-Conv to produce the latent representation.

The decoder uses a similar architecture to match the capacity of the encoder: after a transformation by Conv-ReLU, the vector is split and put to multiple transposed 1-D convolutions with the same kernel sizes as in the encoder. Finally, the estimated signal \(\mathbf{\bar{s}}_i\) is the sum of these outputs.

%% file: 04_experiments.tex
\section{Experiments}

\subsection{Experimental Setup}

For quantitative comparison against existing systems, we use the MUSDB18~\cite{musdb18} dataset consisting of 86 train, 14 validation and 50 test tracks of multi-genre 44.1kHz music.  Each track is annotated with \emph{vocals}, \emph{drums}, \emph{bass}, and \emph{other}.  

We augment the data using standard techniques~\cite{dataAugmentation}: random cut of the 8-second training sample, random amplification of each source from \((0.75,1.25)\), random selection of the left or the right channel and shuffling of the sources between different tracks in half of the training batch.  We train using RAdam~\cite{liu2019radam} with the Lookahead optimizer~\cite{zhang2019lookahead} for a max of 250 iterations. Further hyperparameter settings and training details are available together with the code\footnote{\url{https://github.com/pfnet-research/meta-tasnet}}.

\subsection{Results}

We examine the effectiveness of meta-learning when compared to various other forms of parameter-sharing, and also report the importance of modifications to the Conv-TasNet architecture (Table~\ref{table:ablation}).

\noindent \textbf{Conv-TasNet Modifications} We find that each modification introduced in Sec.~\ref{subsec:improvements} improves performance over vanilla Conv-TasNet, with the multi-stage model resulting in the largest gains (a~{\raise.17ex\hbox{$\scriptstyle\sim$}}.3 increase in average SI-SNR over all instruments).  The combination of all such improvements is denoted as the Baseline system, and improves average SI-SNR by nearly a entire point.  This makes it comparable with other state-of-the-art approaches that instead use spectrogram representations.

\noindent \textbf{Parameter Sharing} Shifting focus to parameter sharing, we compare three different architectures, each with a different approach to sharing.  The Baseline model is independent, training a separate masking model for each instrument.  Finding inspiration from multitask models~\cite{hyperface}, we also experiment with tying the TCN layer parameters, sharing them across instruments. This requires other layers in the masking network to learn instrument-specific projections after the TCN.  Finally we present a meta-learning model (Ours). 

Unfortunately we were not able to train a competitive multi-task model, and our shared TCN performs significantly below our improved Baseline system.  The Baseline is a strong model, and we find it even outperforms meta-learning on two instruments (vocals and drums).  In terms of overall performance, the meta-learning model achieves a modest improvement of SI-SNR by 0.8 averaged across all instruments.

We also compare against previously published and state-of-the-art systems for MUSDB18 (Table~\ref{table:results}). 
Here SDR was either taken from the respective papers (\cite{chandna2017monoaural}\cite{lluis2018end}\cite{open-unmix}) or from the SiSEC18~\cite{SiSEC18} evaluation scores (\cite{stoller2018wave}\cite{liu2018denoising}\cite{takahashi2018mmdenselstm}), and we show the median of frames, median of tracks, evaluated with BSSEval v4~\cite{SiSEC18}. Since our loss is scale-invariant and the BSSEval v4 is scale-dependent, we scale the estimations \(\mathbf{\bar{s}}_i\) by \(\bm{\alpha} = \mathrm{argmin}_{\bm{\alpha}}(\mathbf{s} - \sum_i \bm{\alpha}_i\mathbf{\bar{s}}_i)^2\).

The combined improvements of our proposed modification and meta-learning yields results comparable to state-of-the-art (MMDenseLSTM), and new best scores on \emph{bass} and \emph{other} categories.  Notably the previous best performance for a time domain-based model was from Wave-U-Net, and we improve upon this type of model by a large margin.

Meta-learning in this domain presents interesting trade-offs.  In situations where maximum performance and model size are important (such as on device applications), our meta-learned model achieves slightly higher performance and has 4x fewer\footnote{The encoder has roughly 9.17M parameters, the decoder 3.83M and the parameter generator 32.54M.} masking parameters than our baseline.  If training is constrained in terms of time or memory, our Conv-TasNet Baseline is a model that reaches convergence quickly, and provides approximately state-of-the-art performance.

\begin{table}
\footnotesize
\begin{tabular}{l r r r r r}
\toprule
& \textbf{vocals} & \textbf{drums} & \textbf{bass}  & \textbf{others} & \textbf{avg}\\
\midrule
Conv-TasNet &  5.91 & 6.35 & 4.53 & 2.28 &  4.76\\
\hspace{3 mm} + stronger enc. & 6.32  & 6.17 & 4.60 & 2.50& 4.90\\
\hspace{3 mm} + aux loss & 5.65 & 6.21 & 4.54 & 2.05 & 4.61\\
\hspace{3 mm} + multi-stage & 5.75 & 7.22 & 5.13 & 2.18 & 5.07\\
\midrule
Baseline & \textbf{6.36} & 7.41 & 5.85 & \textbf{2.93} & 5.64\\
Shared TCN & 5.35 & 7.08 & 5.55 & 2.26 & 5.06\\
\midrule
Meta-TasNet & 6.26 & \textbf{7.68} & \textbf{6.11} & 2.86  & \textbf{5.72}\\
\bottomrule
\end{tabular}
\caption{Ablations of our improvements to Conv-TasNet, and comparisons between our method and other alternatives to parameter sharing (Baseline as no sharing).  We report the SI-SNR value on MUSDB18 dev dataset.}
\label{table:ablation}
\end{table}

\begin{table}[t]
\footnotesize
\begin{tabular}{l r r r r r}
\toprule
& \textbf{vocals} & \textbf{drums} & \textbf{bass} & \textbf{other} & \textbf{avg} \\
\midrule
DeepConvSep~\cite{chandna2017monoaural} & 2.37 & 3.14 & 0.17 & -2.13 & 0.89 \\
WaveNet$^{*}$~\cite{lluis2018end} & 3.35 & 4.13 & 2.49 & 0.41 & 2.60 \\
Wave-U-Net$^{*}$~\cite{stoller2018wave} & 3.25 & 4.22 & 3.21 & 2.25 & 3.23 \\
Spect U-Net~\cite{liu2018denoising} & 5.74 & 4.66 & 3.67 & 3.40 & 4.37 \\
Open Unmix~\cite{open-unmix} & 6.32 & 5.73 & 5.23 & 4.02 & 5.36 \\
MMDenseLSTM~\cite{takahashi2018mmdenselstm} & \bf{6.60} & \bf{6.41} & 5.16 & 4.15 & \bf{5.58} \\
\midrule
\textbf{Meta-TasNet$^{*}$} & 6.40 & 5.91 & \bf{5.58} & \bf{4.19} & 5.52 \\
\bottomrule
\end{tabular}
\caption{A comparison SDR scores of our proposed approach with other systems on the test section of MUSDB18 dataset (*indicates the system works directly in the time domain).} 
\label{table:results}
\end{table}

%% file: 05_related.tex
\section{Related Work}

A major design choice in music source separation models is whether to (1) train a separate model for each instrument~\cite{open-unmix}, (2) to use a single class-conditional model, or (3) to use an instrument agnostic approach~\cite{recursive-tasnet}.  Our approach aims to combine the advantages of the first two; the high-precision of independent models, with improved optimization via parameter sharing in single models.  It is also an effort to incorporate prior source knowledge into TasNet-type models.

Parameter sharing is an important design decision in many neural architectures, and many methods exist, ranging from the standard multi-task formulation, to conditioning on an embedded representations in a similar manner as we do (query-based networks~\cite{query-source-sep}), or generating parameters of a mixture model used during mask creation~\cite{class-conditional-source-sep}.  We pursue these goals through other means (end-to-end training parameter generation models), and achieve significantly better performance than their reported results.

In terms of methodology, our work is directly inspired by meta-learning, especially for sequence modeling~\cite{meta-learning-LM}, and where a generator network predicts the parameters of a second network.  This fits the paradigm set forth by HyperNEAT~\cite{hyperNEAT} and hyper-networks~\cite{hyper-networks}.  As in the latter, our architecture is learned end-to-end, making perhaps the largest disadvantage one of training speed: smaller, better extractors at test time come at the cost of a many-fold increase in training time.

Such approaches have also been applied to machine translation (referred to as \emph{context parameter generation}~\cite{platanios-etal-2018-contextual}).  Similarly,
we focus solely on parameter generation, but view learning additional optimization parameters (learning rate, layer architecture) as promising future work.  

%% file: 06_conclusion.tex
\section{Conclusion}

We have shown that meta-learning extractor parameters can yield many benefits, including better performance and smaller model sizes.  In future work we wish to apply this approach to larger, more diverse sets of instruments, where such benefits should be more pronounced. And because instrument envelope durations vary, a fixed-width window for all representation learning is not optimal, and meta-learning more significant architectural choices may yield further improvements. 

\section{Acknowledgements}

We would like to thank Motoki Abe, Huachun Zhu, and Daiki Higurashi for helpful feedback and discussions.

\clearpage